\documentclass[twocolumn,showpacs,preprintnumbers]{revtex4}
\usepackage{amssymb}
\usepackage{amsfonts}
\usepackage{amsmath}
\usepackage{graphicx}
\usepackage{dcolumn}
\usepackage{bm}

\input{tcilatex}

\begin{document}

\title{Influence of relaxation on propagation, storage and retrieving of
light pulses in electromagnetically induced transparency medium}
\author{Gor Nikoghosyan}
\author{Gayane Grigoryan}
\date{\today }

\begin{abstract}
By solving the self-consistent system of Maxwell and density matrix
equations to the first order with respect to nonadiabaticity, we obtain an
analytical solution for the probe pulse propagation. The conditions for
efficient storage of light are analyzed. The necessary conditions for
optical propagation distance has been obtained.
\end{abstract}

\preprint{APS/123-QED}
\affiliation{Institute for Physical Research, 378410, Ashtarak-2, Armenia}
\maketitle





\section{\label{sec:level1}Introduction}

Recent advances in quantum information science have shown that, on one hand,
photons are ideal carriers of quantum information, and on the other hand,
atoms represent reliable and long-lived storage and processing units. In
recent years quantum light storage is one of the extensively studied tasks
of quantum optics. Basic idea of light storage is electromagnetically
induced transparency (EIT) \cite{EIT}.

Electromagnetically induced transparency is a coherent interaction process
in which a coupling laser field is used to made the optical dense media
transparent for the probe field. Since its discovery, a number of new
effects and techniques for light-matter interaction have appeared [2-6].
Most notably, from the point of view of the work presented here, particular
attention has been devoted to ultraslow light propagation and light storage
techniques [3-5].

The key concept of EIT is the dark state and population trapping \cite%
{arrimon}. The dark state is a specific coherent superposition state which
does not contain excited short-living atomic level due to\ destructive
interference between two interaction paths. The dark-state is eigenstate of
the light -- atom interaction Hamiltonian, so the atom prepared in a
dark-state can not be excited and cannot leave the dark-state if the
interaction is adiabatic (Fig.1). The population trapping via applying
strong coupling leads to the adiabatic formation of the dark-state. Since
the interaction is realized by the light pulses the infuence of nonadiabatic
corrections may become important \cite{adiabatica}, \cite{Nasha} . This
infuence has been studied e.g. in \cite{adiabatica}. In particular, the
first nonadiabatic correction connects the dark state and the bright state,
so the depletion of the bright state, because of optical pumping, can affect
the dark state. Despite of large amount of experimental and theoretical
papers concerning light storage (see \cite{revlukin}\ and citations there),
and applications in quantum information science, the influence of
decoherence level width on information carried by stored light is studied
insufficiently.

In this work we present theoretical study which discusses and explains
influence of all relaxations on probe propagation both analytically and
numerically (it is essential in especially, solid state systems \cite{solid}%
). The goal is to study comprehensively how the depletion of bright state
will affect the pulse propagation in an EIT\ media and light storage in
particular.\ By solving the coupled system of Maxwell and density matrix
equations to the first order of the nonstationary pertrubation theory with
respect to nonadiabaticity and decoherence we obtain analytical solution
which completely describes the probe pulse propagation and is consistent
with the recent light storage experiments.

The paper is organized as follows. In section II the basic equations are
written down and the probe pulse propagation equation is derived and
analyzed. In section III and Appendixes the analytical solution for
counterintuitive pulse switching order and for matched pulses are obtained
and the asymptotic solutions discussed. Section IV deals with the physical
consequences of the obtained solution, namely the necessary conditions of
the pulse storage and retrieving, also the numerical results are
demonstrated. In section V we consider the transverse relaxation of the
coherence induced in the medium. Section VI concludes the paper.

\section{\label{basic}Basic Equations}

Figure 1 shows a schematic diagram of the atomic system in the EIT basis:
media of three level atoms interacting with two laser pulses $%
E_{p}=A_{p}\cos \left( \omega _{p}t-k_{p}z+\varphi _{p}\right) $ (probe)\
and $E_{c}=A_{c}\cos \left( \omega _{c}t-k_{c}z+\varphi _{c}\right) $\emph{\ 
}(coupling). The probe field resonantly connects the state $|1\rangle $ to
the state $|3\rangle $ and the coupling field connects $|2\rangle $ to $%
|3\rangle $. The Hamiltonian of the system in the rotating wave
approximation is: 
\begin{equation*}
H=\hbar \Delta \sigma _{33}-\hbar \Omega _{p}\sigma _{31}-\hbar \Omega
_{p}\sigma _{32}+H.c.\text{,}
\end{equation*}%
where $\Omega _{p,c}=\dfrac{A_{p,c}\mu _{3i}}{\hbar }$ ($i=1,2$) are the
respective Rabi frequencies, $\sigma _{ij}=|i\rangle \langle j|$ are the
atomic transition operators, $\Delta =\omega _{p}-\omega _{31}=$ $\omega
_{c}-\omega _{32}$ is the detuning of the pulse frequencies from the upper
level and $\mu _{3i}$ ($i=1,2$) are the dipole moments of corresponding
transitions.

We assume\ that: (i) the probe field is weak as compared to the coupling
pulse field $\Omega _{p}<<\Omega _{c}$; (ii) the interaction is adiabatic ($%
\Omega _{c}T>>1$, where $T$ is the interaction duration). Then, the atomic
density matrix equation may be written as

\begin{eqnarray}
\overset{.}{\rho }_{31} &=&-\Gamma \rho _{31}+i\Omega _{p}+i\Omega _{c}\rho
_{21},  \notag \\
\overset{.}{\rho }_{21} &=&i\Omega _{c}^{\ast }\rho _{31},  \label{bloch} \\
\rho _{11} &=&1,  \notag \\
\rho _{22} &=&\rho _{33}=\rho _{32}=0,  \notag
\end{eqnarray}%
where $\Gamma $ is the width of the upper level which is the sum of the
spontaneous decay and transverse relaxations rates. It is supposed that
interaction is fast enough to neglect the decoherence between metastable
levels (sec. III, IV), or to take it into account to the first order (sec.
V).

The propagation of the pulses is governed by the Maxwell equation for slowly
varying amplitudes,%
\begin{eqnarray}
\left( \frac{\partial }{\partial x^{\prime }}+\frac{1}{c}\frac{\partial }{%
\partial t^{\prime }}\right) \Omega _{p} &=&iq_{p}\rho _{31},  \label{prop}
\\
\text{ \ }\left( \frac{\partial }{\partial x^{\prime }}+\frac{1}{c}\frac{%
\partial }{\partial t^{\prime }}\right) \Omega _{p} &=&iq_{c}\rho _{32}, 
\notag
\end{eqnarray}%
where $q_{i}=\dfrac{2\pi \mu _{3i}\omega _{i}N}{\hbar c}$, $N$ is the atomic
number density.

System of equations (\ref{bloch}) can be reduced to one equation for $\rho
_{31}$%
\begin{equation}
\overset{..}{\rho }_{31}-\overset{.}{\rho }_{31}\frac{\overset{.}{\Omega }%
_{c}}{\Omega _{c}}+\rho _{31}\Omega _{c}^{2}+\Gamma \left( \overset{.}{\rho }%
_{31}-\rho _{31}\frac{\overset{.}{\Omega }_{c}}{\Omega _{c}}\right) =i\Omega
_{c}\overset{.}{\theta }  \label{dens}
\end{equation}%
where $\theta =\dfrac{\Omega _{p}}{\Omega _{c}}$ is the common used notation
for the so called mixing angle. Influence of the first two terms in (\ref%
{dens}) can be neglected if we confine to\ only first terms with respect to
the nonadiabaticity (i.e. $\left( \Omega _{c}T\right) ^{-2}<<1$ is
neglected). Influence of the fourth term in (\ref{dens}) is essential
parameter only under the assumption%
\begin{equation}
\Gamma T>>1.  \label{largeg}
\end{equation}%
The meaning of the condition (\ref{largeg})\ is obvious: under the condition
of complete adiabaticity relaxation does not affect the pulse propagation
(dark-state), but taking into account first nonadiabatic correction, has
essential influence. The relaxation can be neglected when $\Gamma T\lesssim
1 $.

Finally, by substituting the Maxwell equation (\ref{prop}) into (\ref{dens})
one gets pulse propagation equation in wave variables $x=x^{\prime }$, $%
t=t-x^{\prime }/c$,%
\begin{equation}
\frac{1}{\Gamma _{1}}\frac{\partial ^{2}\theta }{\partial x\partial t}+\frac{%
q_{p}}{\Omega _{c}^{2}}\frac{\partial \theta }{\partial t}+\frac{\partial
\theta }{\partial x}=0,  \label{probe_prop}
\end{equation}

where notation $\Gamma _{1}=\dfrac{\Omega _{c}^{2}}{\Gamma }$ is used. In
this connection the coherence dynamics is governed by the following equation:%
\begin{equation}
\overset{.}{\rho }_{21}=-\Gamma _{1}\rho _{21}-\Gamma _{1}\theta
\label{coh_dyn}
\end{equation}

Thus $\Gamma _{1}$ is the coherence decay rate, or width of EIT\ resonance,
due to applied coupling. Under the condition%
\begin{equation}
\Gamma _{1}T>>1  \label{bright}
\end{equation}%
equation (\ref{coh_dyn}) has the well known quasi-stationary solution $\rho
_{21}=-\theta $ \cite{polariton}, and the equation (\ref{probe_prop}) passes
to the dark state polariton propagation equation. The condition (\ref{bright}%
) means, that width of EIT\ resonance exceeds the spectral width of the
probe.

\section{\label{solution of propag}Solution of propagation equation}

The obtained probe pulse propagation equation (\ref{probe_prop}) is solved
by the method presented in \cite{Mostowski}. Since (\ref{probe_prop}) is
linear in $\theta $ and $\Omega _{c}\left( t\right) $ is independent of $x$,
it can be solved by using the Laplace transform with respect to $x$. The
solution of (\ref{probe_prop}) for $\theta $'s Laplace image can be found
easily:%
\begin{equation}
\overset{\symbol{126}}{\theta }\left( s,t\right) =\int\limits_{-\infty }^{t}%
\frac{\overset{.}{\theta }_{0}+\Gamma _{1}\theta _{0}}{s+q_{p}/\Gamma }%
B\left( s,t,t_{1}\right) dt_{1}+c\left( s\right) B\left( s,t,-\infty \right)
\label{image_sol}
\end{equation}%
where $B\left( s,t,t_{1}\right) =\exp \left( -\dfrac{s}{s+q_{p}/\Gamma }%
\int\limits_{t_{1}}^{t}\Gamma _{1}dt^{\prime }\right) $, $c(s)$ is an
integration constant that is determined by the initial condition, $c\left(
s\right) =\overset{\symbol{126}}{\theta }\left( s,-\infty \right) $. If
pulses are switched in counterintuitive sequence (coupling turns on earlier
than the probe does) then $c\left( s\right) =0$, since $\theta \left(
z,-\infty \right) =\theta _{0}\left( -\infty \right) =0$ (see appendix A).

Space time evolution of the probe pulse is obtained by implementing the
reverse Laplace transform in (\ref{image_sol}).%
\begin{eqnarray}
\theta \left( z,t\right) &=&\int\limits_{-\infty }^{t}dt_{1}\left( \theta
_{0}\left( t\right) \Gamma _{1}+\overset{.}{\theta }_{0}\left( t\right)
\right) \times  \label{solution} \\
&&\times \exp \left( -z-\alpha \left( t_{1},t\right) \right) I_{0}\left( 2%
\sqrt{z\alpha \left( t_{1},t\right) }\right) ,  \notag
\end{eqnarray}

where $z=\dfrac{q_{p}x}{\Gamma }$ is propagation distance normalized to
linear absorption factor and, for convenience, the notation $\alpha \left(
t_{1},t\right) =\int\limits_{t_{1}}^{t}\Gamma _{1}\left( t^{\prime }\right)
dt^{\prime }$ is used. \ By using the condition $z\alpha \left(
t_{1},t\right) >>1$ one can substitute the modified Bessel function by its
asymptote, so the solution (\ref{solution})\ reduces to the following:

\begin{eqnarray}
\theta \left( z,t\right) &=&\frac{1}{2\sqrt{\pi }}\int\limits_{-\infty
}^{t}dt_{1}\left( \theta _{0}\left( t\right) \Gamma _{1}\left( t_{1}\right) +%
\overset{.}{\theta }_{0}\left( t\right) \right) \times  \label{gauss} \\
&&\times \exp \left( -\left( \sqrt{z}-\sqrt{\alpha \left( t_{1},t\right) }%
\right) ^{2}\right) \left( z\alpha \left( t_{1},t\right) \right) ^{-1/4}. 
\notag
\end{eqnarray}

Depending on optical propagation distance $z$ two simple asymptotes for (\ref%
{gauss}) can be obtained (see appendix B). The first is the case where%
\begin{equation}
\frac{\Gamma _{1m}T}{\sqrt{z}}>>4\sqrt{\ln 2}  \label{polariton_cond}
\end{equation}%
$\Gamma _{1m}$ is the maximal value of $\Gamma _{1}\left( t\right) $ (see
also \cite{polariton}). Under condition (\ref{polariton_cond}), solution (%
\ref{solution}) reduces to the dark-state polariton propagation solution
with correction in (\ref{bright}):%
\begin{equation}
\theta \left( z,t\right) =\theta _{0}\left( \xi \right) +\frac{1}{\Gamma
_{1}\left( \xi \right) }\overset{.}{\theta }_{0}\left( \xi \right)
\label{polariton}
\end{equation}%
where $\xi $ is the non-linear time determined by $\int\limits_{\xi
}^{t}\Omega _{c}^{2}\left( t_{1}\right) dt_{1}=q_{p}x$ \cite{Nasha}. Note,
that turning off the coupling $\Omega _{c}\left( t\right) $ does not reduce $%
\Gamma _{1}\left( \xi \right) $\ to zero, since $\xi $ retards from $t$. In
this case, as it will be shown below, the information stored in the medium
can be well retrieved (see sec. IV).

In the case of condition reversed to (\ref{polariton_cond}),%
\begin{equation}
\frac{\Gamma _{1m}T}{\sqrt{z}}<<4\sqrt{\ln 2}.  \label{bluring_cond}
\end{equation}%
the\emph{\ }solution (\ref{solution}) reduces to 
\begin{equation}
\theta \left( z,t\right) =R\exp \left( -\left( \sqrt{z}-\sqrt{\alpha \left(
t_{0},t\right) }\right) ^{2}\right) \left( z\alpha \left( t_{0},t\right)
\right) ^{-1/4}  \label{bluring}
\end{equation}%
where $t_{0}$ is the maximal value of $\theta _{0}\left( t\right) $, $%
R=\int\limits_{-\infty }^{\infty }\Gamma _{1}\left( t^{\prime }\right)
\theta _{0}\left( t^{\prime }\right) dt^{\prime }$ and does not depend on
time. We emphasize that for propagation distances meeting the condition (\ref%
{bluring_cond}), the obtained pulse loses all the information about its
initial temporal shape, since the right hand side in (\ref{bluring}) does
not contain time dependent $\theta _{0}$.

\section{\label{calc}Discussion}

In this section the probe pulse propagation dynamics obtained from the
analytical solution (\ref{solution}) is presented. First of all we consider
the case of constant coupling field. Shape of the initial pulse is chosen to
be double-humped in order to visualize the propagation dynamics. For the
small propagation distances when the condition (\ref{polariton_cond})\ is
met influence of $\Gamma $ is negligible (Fig.2a). When the condition (\ref%
{polariton_cond}) is violated, the influence of upper level width becomes
essential as one can see from Figs. 2b,c. Thus influence of $\Gamma $ breaks
the adiabaton propagation regime.

As it was mentioned above, propagation over very long distances (\ref%
{bluring_cond}) leads to the lost of the information on the initial pulse
temporal shape. This can be seen in Fig 3, where propagation over the same
distance of two pulses with different temporal shapes but with the same
initial area is depicted. By propagating over very long distance (\ref%
{bluring_cond}) they lose any information about their initial temporal
shapes.

In Fig. 4 we show that the increase of $\Gamma _{1}$ suppresses the smearing
of the probe. This is caused by the decrease of the bright state population
and hence leads to the decrease the influence of $\Gamma $ on pulse
propagation. Note, that in the dark-state propagation regime the pulse
temporal shape does not depend on coupling field amplitude or on unstable
level width.

Summarizing presented results one can see that to minimize the pulse
smearing during its propagation one has to either increase $\Gamma _{1}$ or
decrease the propagation distance $z$. The situation changes dramatically
for the light storage and retrieving process ($\Omega _{c}\neq const$).

It is known, that pulse can be completely stored and retrieved from the
medium if the medium length and $\Gamma _{1}$ meet the condition (see for
example \cite{Nasha}):%
\begin{equation}
z\gtrsim \Gamma _{1m}T.  \label{fitting}
\end{equation}

For efficient storage and retrieving the condition (\ref{polariton_cond})
also has to be met. Combining this two nonequalities one gets that to
completely store and well retrieve the light pulse, $\Gamma _{1}$ has to
meet the following condition:$\ $%
\begin{equation}
\Gamma _{1m}T>>16\ln 2>>1.  \label{good_storage}
\end{equation}

Therefore, influence of the second term in (\ref{polariton}) is insufficient
when the condition (\ref{good_storage}) is met.

Storage and retrieving of the light pulse for different propagation
distances under the condition (\ref{good_storage}) is depicted in Fig 5. For
small propagation distances when the condition (\ref{fitting}) is violated
only the falling edge of the pulse is stored and can be retrieved (Fig. 5a),
because when this edge enters the medium, the leading edge emerges already.
For larger propagation distances when the condition (\ref{fitting}) is
satisfied the whole pulse can be stored and then well retrieved by turning
on the coupling field.

Let us now consider the case when the condition (\ref{good_storage}) is not
met (Fig. 6). For small propagation distances when the condition (\ref%
{polariton_cond}) is satisfied but (\ref{fitting})\ is not, only the falling
edge of the pulse can be stored and retrieved. Propagation over longer
distances brings to satisfying of (\ref{fitting})\ and violation of (\ref%
{polariton_cond}). Thus the whole pulse can be stored but the retrieved
pulse temporal shape is smeared.

We present finally comparison of the experimental results with our
analytical solution. In Fig 7a the experimental data of storage and
retrieving of light pulse from \cite{experiment} are presented. Curve in Fig
7b is plotted from our analytical solution (\ref{solution}): all parameters
correspond to the conditions of the experiment. One can see good consistency
between experimental data and our analytical solution (Note that the storage
in case of experiment () is incomplete as was discussed above).

\section{\label{sec_V}Consideration of $\protect\rho _{21}$ transverse decay}

In this sections we take into acount the quantity $\gamma T$ in first order.
This leads, instead of (\ref{bloch}), to the equations.

\begin{eqnarray*}
\overset{.}{\rho }_{31} &=&-\Gamma \rho _{31}+i\Omega _{p}+i\Omega _{c}\rho
_{21}, \\
\overset{.}{\rho }_{21} &=&-\gamma \rho _{21}+i\Omega _{c}^{\ast }\rho _{31},
\\
\rho _{11} &=&1, \\
\rho _{22} &=&\rho _{33}=\rho _{32}=0,
\end{eqnarray*}
Thus, probe pulse propagation equation is written as follows:%
\begin{equation}
\frac{1}{\Gamma _{1}}\frac{\partial ^{2}\theta }{\partial x\partial t}+\frac{%
\partial \theta }{\partial x}+\frac{q_{p}}{\Gamma _{1}\left( \Gamma +\gamma
\right) }\frac{\partial \theta }{\partial t}+\frac{q_{p}\gamma }{\Gamma
_{1}\left( \Gamma +\gamma \right) }\theta =0,  \label{V_2}
\end{equation}

where $\Gamma _{1}\left( t\right) $ now is%
\begin{equation}
\Gamma _{1}\left( t\right) =\frac{\Omega _{c}^{2}+\gamma \left( \Gamma +%
\dfrac{\dot{\Omega}_{c}}{\Omega _{c}}\right) }{\Gamma +\gamma }.  \label{V_3}
\end{equation}

As results from (\ref{V_3}), to completely stop the light in the medium $%
\left( \Gamma _{1}=0\right) $ one should turn off the coupling field $\Omega
_{c}$ at the rate $\Gamma $ (i.e., $\Gamma +\dfrac{\dot{\Omega}_{c}}{\Omega
_{c}}=0$).

By performing the stated above Laplace transform procedure one obtains
analytical solution of the equation (\ref{V_2}) in the form%
\begin{eqnarray}
\theta \left( z,t\right) &=&\int\limits_{-\infty }^{t}dt_{1}\left( \theta
_{0}+\Gamma _{1}\dot{\theta}_{0}\right) \times  \label{V_4} \\
&&\times \exp \left( -z-\int\limits_{t_{1}}^{t}\Gamma _{1}dt^{\prime
}\right) I_{0}\left( 2\sqrt{z\alpha \left( t_{1},t\right) }\right)  \notag
\end{eqnarray}%
where $\alpha \left( t_{1},t\right) =\int\limits_{t_{1}}^{t}\Gamma
_{1}\left( t^{\prime }\right) -\gamma dt^{\prime }$ and $z=\dfrac{q_{p}x}{%
\Gamma +\gamma }$.

More detailed analysis of the expression (\ref{V_2}) will be performed in a
subsequent publication.

\section{Conclusion}

We have considered the propagation, storage and retrieving of the light
pulse in EIT media by taking into account all dephasing rates. From coupled
system of Maxwell and density matrix equations we derive the probe pulse
propagation equation, which in particular case passes into the dark-state
polariton propagation equation. We find an analytical solution and analyzed
its physical consequences. We derived a simple asymptotes of the solution,
and showed strong dependence of light pulse temporal shape on optical
propagation distance in the presence of relaxations. We demonstrated that an
efficient storage of light is possible by choosing appropriate coupling
intensities and optical propagation distances. Finally, we compared our
solution with experimental data and showed that our solution is well
consistent with the recent experiments.

\begin{acknowledgments}
We are grateful to Prof. M.Fleischhauer, Prof. V.Chaltykyan and Prof.
Yu.Malakyan for helpful discussions. The work was supported by the ISTC
Grant No. \#A-1095.
\end{acknowledgments}

\bigskip

\appendix

\section{ }

To find the $c\left( s\right) $ in general case we examine the equation (\ref%
{prop}).

From this equation one obtains $\dfrac{\partial \theta }{\partial x}=\dfrac{%
iq_{p}}{\Omega _{c}}\rho _{31}\left( x,t\right) $, and the solution can
formally be written as $\theta \left( x,-\infty \right) =\dfrac{iq}{\Omega
_{c}}\int\limits_{0}^{x}\rho _{31}\left( x^{\prime },-\infty \right)
dx^{\prime }+\theta _{0}\left( -\infty \right) \,$. This can be simplified
by taking into account that there is no dipole moment induced before the
interaction is turned on ($\rho _{31}\left( x,-\infty \right) =0$),%
\begin{equation*}
\theta \left( x,-\infty \right) =\theta _{0}\left( -\infty \right)
\end{equation*}

For the counterintuitive order of pulse switching we have $\theta _{0}\left(
-\infty \right) =0$, hence $c\left( s\right) =0$. In the case of
simultaneous pulse switching $\theta _{0}\left( -\infty \right) =const$,
therefore $c\left( s\right) =\dfrac{const}{s}$.

In the case $\theta _{0}=const$ from general solution (\ref{image_sol}) one
easily obtains the so-called matched pulse propagation regime $\theta
=\theta _{0}$ \cite{match}.

\section{ }

The integrand function in (\ref{gauss}) is the product of two time dependent
bell-shaped functions. This kind of integral is evaluated easily when the
temporal width of one function is much larger than the width of the other
one. Then by the first order saddle-point technique the narrow function can
be approximated by delta function.

The time widths of the first and second functions in (\ref{gauss}) are,
respectively, the interaction duration $T$ and $\dfrac{4\sqrt{z\ln \left(
2\right) }}{\Gamma _{1m}}$

On the one hand, if the condition (\ref{polariton_cond}) is met, the
integration of (\ref{gauss})\ gives the well known dark-state polariton
solution (\ref{polariton}). On the other hand, if the condition (\ref%
{bluring_cond}) is satisfied, integration of (\ref{gauss})\ by the
saddle-point technique brings to the asymptote (\ref{bluring}).

\bigskip

Fig.1 Schematic diagram of the three-level atomic system in the basis of EIT
in the dressed-state (a) and the bare-state (b) representations. The field $%
\Omega _{c}$ couples the $\left\vert bright\right\rangle $ state to the
upper level $\left\vert 3\right\rangle $, and another field with Rabi
frequency $\overset{.}{\theta }$ couples the $\left\vert dark\right\rangle $
and $\left\vert bright\right\rangle $ states.

Fig.2 Probe pulse temporal shape after propagation over different
propagation distances in the case of constant coupling field; $\Gamma
_{1}T=40$. Initial temporal shape of the probe has been chosen as $\Omega
_{p}/\Gamma =0.012\exp \left( -\left( \Gamma t/100-7.5\right) ^{2}\right)
+0.01\exp \left( -\left( \Gamma t/100-10\right) ^{2}\right) $. For small
propagation distances when the condition (\ref{polariton_cond}) is met the
influence of upper level width is small (a). Violation of the condition (\ref%
{polariton_cond})\ increases the role of the upper level width: (b), (c).
Here and below the dotted curve is the initial temporal shape of the probe.

Fig.3 Propagation of two probe pulses with different temporal\ shapes but
with the same area over very long distances (\ref{bluring_cond}) in the case
of constant coupling $\Gamma _{1}T=4$. One can see that during propagation
pulse loses all information about its initial temporal shape. The shape of
the probe in the case (a) is the same as in Fig.2, in the case (b) it has
been chosen as $\Omega _{p}/\Gamma =0.022\exp \left( -\left( \Gamma
t/100-8.5\right) ^{2}\right) $.

Fig.4 Probe pulse temporal shape after propagation over the same optical
distance $z=20$ for different constant coupling amplitudes. Increasing of
coupling amplitude decreases the influence of upper level width. The shape
of the probe is the same as in Fig.2.

Fig.5 Storage and retrieving of the probe for different propagation
distances, when the condition (\ref{good_storage}) is met. The coupling
field is chosen as $\Omega _{c}/\Gamma =f\left( t\right) $ \ (a), where $%
f\left( t\right) =1$ for $\Gamma t<1000$ and $\Gamma t>2500$; $f\left(
t\right) =\exp \left( -\left( \Gamma t/100-10\right) ^{2}\right) +\exp
\left( -\left( \Gamma t/100-20\right) ^{2}\right) $ for $1000\leqslant
\Gamma t\leqslant 2500$. For small optical propagation distances only the
falling edge of the pulse is stored and retrieved (b). For larger $z$ the
whole pulse can be stored and well retrieved (c). The shape of the probe is
the same as in Fig.2.

Fig.6 Storage and retrieving of the probe for different propagation
distances when the condition (\ref{good_storage}) is violated. The coupling
field has been chosen as $\Omega _{c}/\Gamma =f\left( t\right) /\sqrt{10}$, $%
\Gamma _{1m}T=40$ (a). Either only the falling edge can be stored and well
retrieved (b) or the whole pulse is stored but loses its initial shape (c).
The shape of the probe is the same as in Fig.2.

Fig.7 Experimental data from \cite{experiment}, (a) and our analytical
solution for parameters corresponding to the experimental conditions (b). $%
\Omega _{p}/\Gamma =\exp \left( -\left( \Gamma t/400-10\right) ^{2}\right) $
for $\Gamma t<1000$, and $\Omega _{p}/\Gamma =\exp \left( -\left( 3\Gamma
t/100-10\right) ^{2}\right) $ for $\Gamma t>1000$. $\Omega _{c}/\Gamma
=0.1f\left( \Gamma t\right) \,$, $\Gamma _{1m}T=5$, time axes is plotted in $%
\mu \sec .$

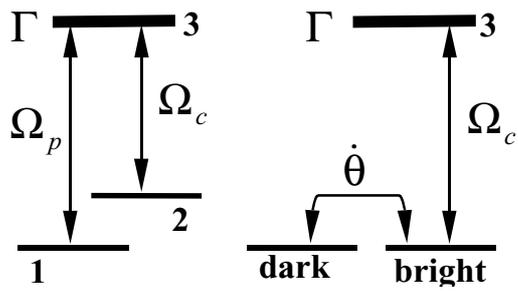
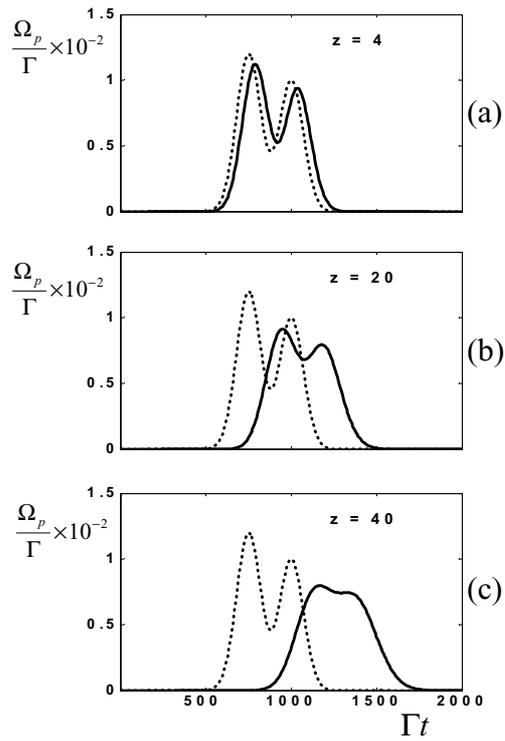
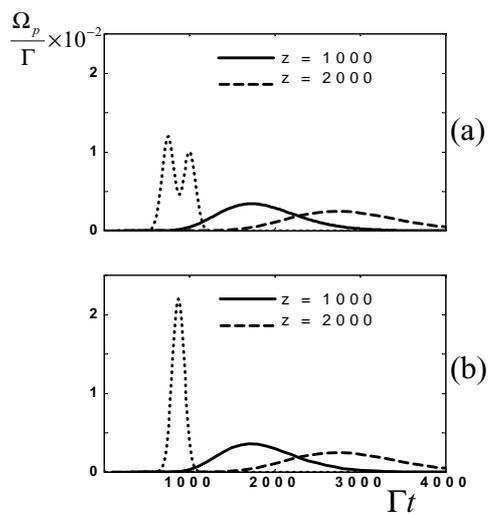
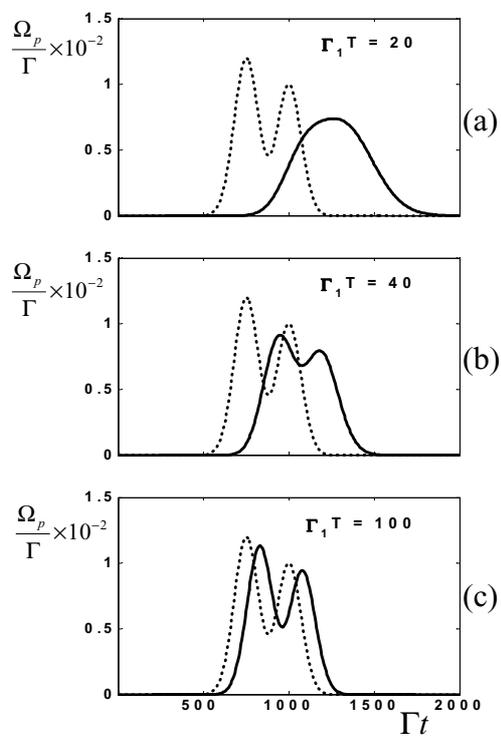
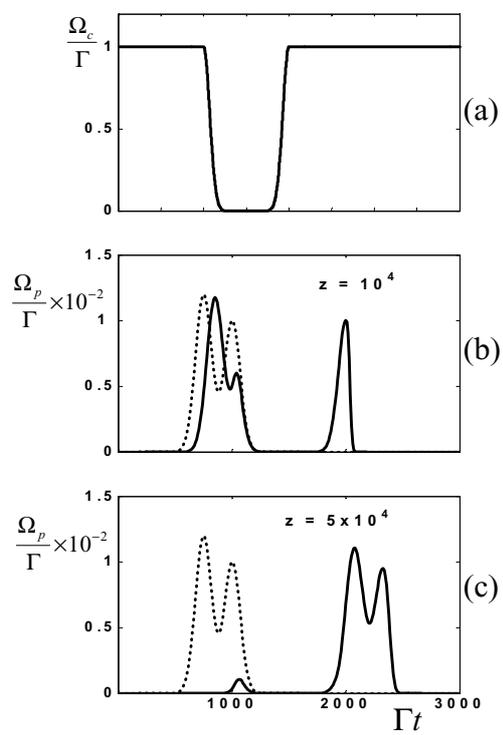
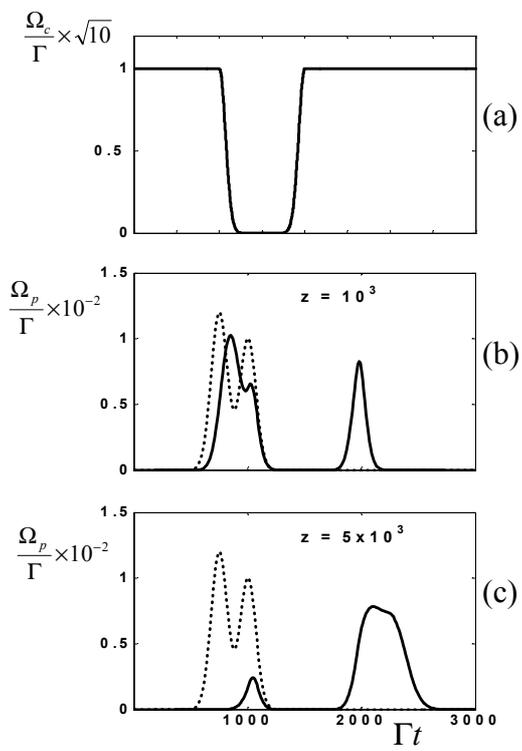
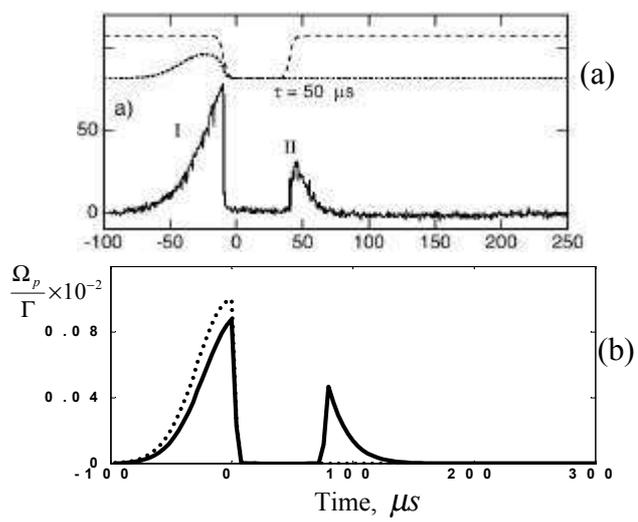
\end{document}